\documentclass[letterpaper,twocolumn,10pt]{article}
\usepackage[twoside=true, head=13pt,
     paperwidth=8.5in, paperheight=11in,
     includeheadfoot=false, columnsep=2pc,
     top=1in, bottom=1in, inner=0.75in, outer=0.75in,
     marginparwidth=2pc,heightrounded]{geometry}
\usepackage[small,compact]{titlesec}
\usepackage{graphicx}
\usepackage[tt=false, type1=true]{libertine}
\usepackage{mathptmx}
\usepackage{amssymb}
\usepackage[varqu]{zi4}
\usepackage[T1]{fontenc}
\usepackage{enumitem}
\usepackage[font={footnotesize},labelfont={footnotesize,bf},textfont={footnotesize,it},skip=4pt]{caption}
\usepackage{amsmath}
\usepackage{booktabs}
\usepackage{tabularx}
\usepackage{cuted}

\newcolumntype{L}{>{\raggedright\arraybackslash}X}
\usepackage{url}
\usepackage[section]{placeins}   
\usepackage[hidelinks]{hyperref}
\usepackage{xurl}
\hypersetup{pdftitle={Disjoint in Name Only: Illusory Resilience in
the Internet}, pdfauthor={}, pdfsubject={}, pdfkeywords={},
pdfcreator={}, pdfproducer={}}
\graphicspath{{fig/}}

\title{Disjoint in Name Only: Illusory Resilience in the Internet}
\author{Subhadeep Sahoo \and Hrishikesh B Acharya}
\date{}

\begin{document}
\pagestyle{plain}
\maketitle
\begin{abstract}
Internet path diversity is usually verified at the logical layer: two
alternatives may share no AS or IP link yet traverse the same facility,
conduit, or submarine landing zone. We call this gap \emph{illusory
resilience} and define the Illusory Resilience Index (IRI) as the latent
hazard-weighted physical overlap of a logically disjoint path pair. To
detect such hidden exposure without carrier underlay maps, we audit three
public signals: correlated BGP withdrawals during documented physical
incidents, operator interconnection-footprint overlap, and shared
submarine landing zones. Two signals survive calibration; the 60-second
BGP signal does not. Across 47 operators in ten regions, same-region
pairs have substantially greater listed facility overlap than
cross-region pairs, and one Marseille campus colocates 153 operator
pairs. Across 68 cable systems in twelve corridors, 31--100\% of
candidate within-corridor pairs share a 25\,km coastal hazard zone; two
corridors are fully saturated, and the same rule recovers ten of eleven
cable pairs documented as co-damaged in 2024. The rejected temporal
predicate has a 0.444 null firing rate and no hits among twelve monitored
positive pairs. Public data therefore expose substantial shared-risk
structure behind nominal diversity and identify where stronger
physical-path verification is warranted.
\end{abstract}


\section{Introduction}
\label{sec:intro}

\begin{figure}[htbp]
\centering
\includegraphics[width=0.98\columnwidth]{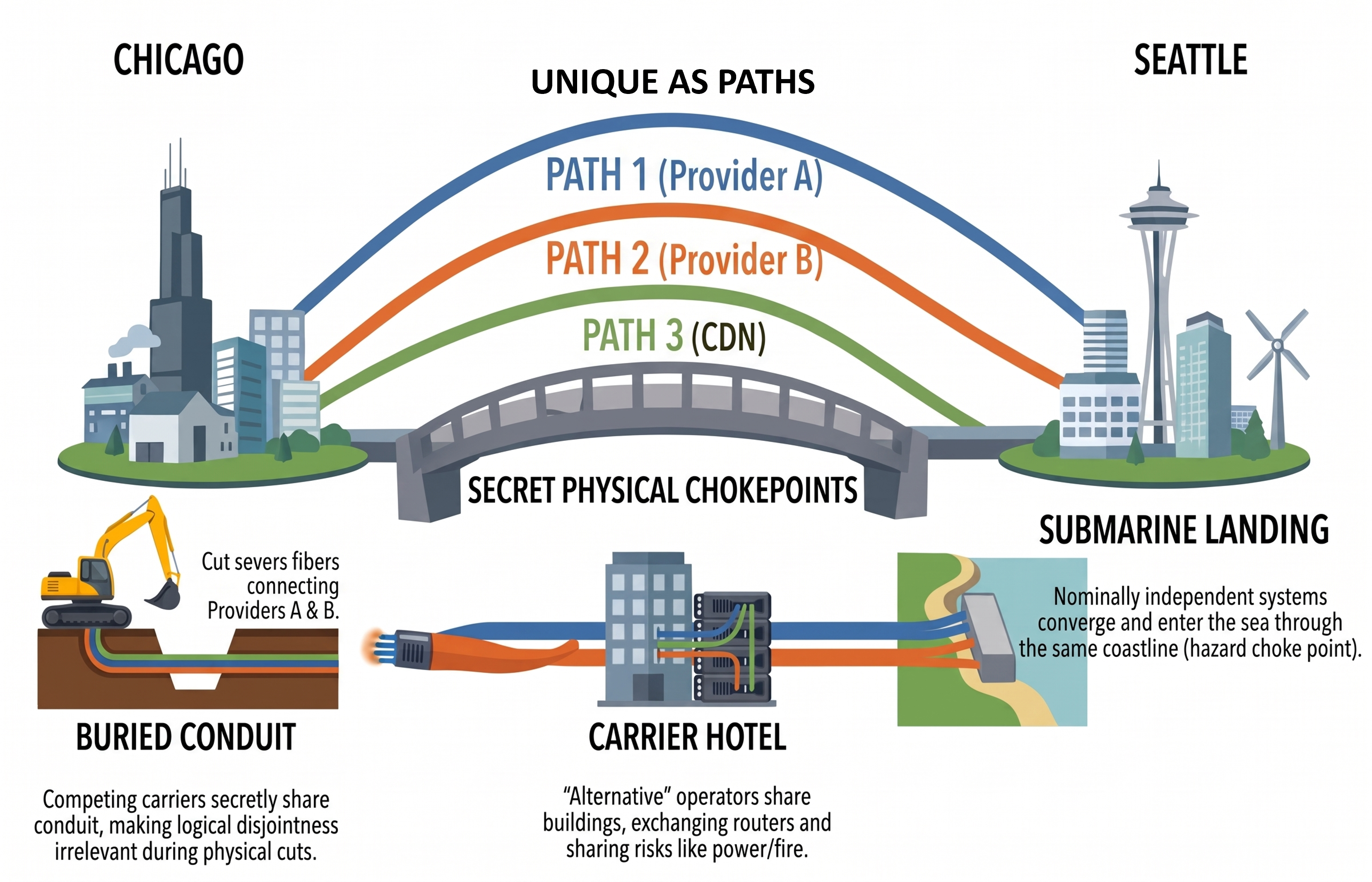}
\caption{The anatomy of illusory resilience.}
\label{fig:intro}
\end{figure}

On an ordinary morning, an excavator outside Chicago severs a buried
conduit. Within the same minute, a company two thousand kilometers away
loses both its links to Seattle: the primary \emph{and} the backup it
bought precisely for this moment (a different provider with a disjoint
AS path and no IP links in common: Fig.~\ref{fig:intro}). Common
logical-layer diversity checks treat the two paths as independent. They
cannot see the shared trench when both providers lease fiber in the same
conduit. In that case, purchasing logical diversity does nothing to
reduce the common physical risk.

The same failure mode appears at several scales. Long-haul fiber from
competing carriers can share rights-of-way; alternative operators can
occupy the same carrier hotels; and distinct submarine systems can enter
the sea through the same few kilometers of coastline
(Fig.~\ref{fig:intro}). We call the gap between logical disjointness and
shared physical exposure \emph{illusory resilience}. The logical path is
visible; the conduits, facilities, landing dependencies, and other
resources that carry it are often proprietary.

Modern cross-layer cartography exposes much more of this hidden underlay.
Public records can reconstruct long-haul fiber maps~\cite{intertubes},
link physical and logical infrastructure~\cite{igdb}, map traceroutes to
submarine cables~\cite{nautilus,calypso}, and combine such maps with
hazard models~\cite{xaminer}. But complete path-level resource sets
remain unavailable outside the carriers, and reconstructed maps inherit
coverage and inference errors from the steps used to build them. We ask a
different, operator-facing question:

\textbf{\emph{Where can public observations expose shared physical risk
behind nominally diverse alternatives, and how strongly does each
observation support that claim?}}

We answer this with a public-data \emph{audit}. Each signal has a defined
native object, a falsifiable firing predicate, and a measured null rate
under a control construction. We evaluate three signals on live data:
correlated BGP withdrawals during documented physical incidents
(temporal), operator interconnection-footprint overlap (structural), and
shared submarine landing zones (geographic). Two survive calibration.
The natural 60-second BGP predicate does not, so it is excluded from the
cross-channel results.

The measurements reveal substantial concentration. Across 1{,}081
operator pairs, the median same-region pair has nonzero listed
facility-footprint overlap, while the median cross-region pair
has none; a single Marseille campus colocates 153 operator pairs. Across
68 cable systems in twelve intercontinental corridors, 31--100\% of
candidate within-corridor pairs enter a common 25\,km coastal hazard
zone, with two corridors fully saturated. The same 25\,km rule recovers
ten of eleven cable pairs documented as damaged together in 2024. The
BGP replay yields a negative result: same-minute co-withdrawal has
a 0.444 null firing rate and no hits among twelve monitored positive
pairs, so we reject it at this granularity.

Our contributions are fourfold. \emph{(1) Formulation:} we define
illusory resilience and the Illusory Resilience Index (IRI), the latent
physical overlap of a logically disjoint path pair
(Sec.~\ref{sec:model}). \emph{(2) Audit method:} we distinguish that
path-level target from the operator, cable-system, and routing objects
that public data actually observe, and pair each signal with an explicit
predicate and control construction (Sec.~\ref{sec:methodology}).
\emph{(3) Measurements:} we quantify operator interconnection
concentration and submarine landing convergence on geographically
diverse live snapshots, and show that the 60-second BGP signal fails
calibration (Sec.~\ref{sec:results}). \emph{(4) Practice:} we produce
traceable cross-layer dossiers that identify where carrier attestation
or path-level verification is most useful, and release the analysis
pipelines and per-figure extracts (Sec.~\ref{sec:leading}). IRI names
physical overlap; the audit measures public evidence of that overlap.

\section{Background and Related Work}
\label{sec:related}

Physical resilience has been studied from several directions, but the
relevant work is split across communities and administrative boundaries. The closest work falls into five groups.

\textbf{Logical resilience and physical failure.}
Foundational Internet measurements established substantial inter-domain
route instability~\cite{paxson96,govindan97} and the importance of
physical causes---fiber cuts, power failures, and hardware faults---in
wide-area outages~\cite{labovitz99}. Geographic studies then showed how
commercial infrastructure can reconcentrate into hub-and-spoke
forms~\cite{grubesic03,grubesic06}; the 2001 Howard Street Tunnel fire is
a canonical example in which several carriers shared one physical
failure domain~\cite{intertubes}. Logical-layer remedies such as
resilient overlays~\cite{ron01}, route repair~\cite{lifeguard12}, and
multihoming reduce many everyday failures, but they cannot certify if
a backup avoids the primary path's physical dependencies.

\textbf{Known physical risk inside a carrier.}
The optical community handles shared physical risk directly when the
plant is known. Shared Risk Link Groups~\cite{srlg}, survivable WDM
routing~\cite{ou05}, and survivable embedding~\cite{narulatam04} all
reason about common physical failure domains. Geographic-failure models
likewise study correlated regional damage~\cite{neumayer11,agarwal11},
and resilience frameworks such as ResiliNets~\cite{sterbenz10,smith11}
and RECODIS~\cite{recodis,gomes16} systematize design under such failure
scenarios. An external customer usually lacks the plant-level
information these methods assume because it stops at the administrative
boundary.

\textbf{Cross-layer cartography and hazard analysis.}
Public-data mapping has steadily reduced that information gap. Internet
Atlas and InterTubes assemble physical maps from public records and
quantify conduit sharing~\cite{atlas13,intertubes}; iGDB links physical
and logical infrastructure along inferred rights-of-way~\cite{igdb};
Nautilus maps IP links to submarine cables~\cite{nautilus}; and Calypso
maps traceroute paths to cable systems~\cite{calypso}. Hazard studies
apply climate~\cite{lightsout18}, earthquake~\cite{mayer21}, solar
storm~\cite{jyothi21}, and disaster-routing~\cite{riskroute} models to
physical infrastructure. Xaminer combines a cross-layer map with a
hazard model to generate risk profiles~\cite{xaminer}; earlier work also
quantified country-level submarine dependence~\cite{bischof18}.

A path-level diversity verdict derived from these maps inherits the
inference chain used to construct them. Traceroute, alias resolution,
ownership attribution, geolocation, and conduit attribution all have
known error modes: alias resolution has a substantial error
literature~\cite{apar09,midar13}, and those errors can change topology
study conclusions~\cite{gunes07}; infrastructure geolocation is
particularly difficult~\cite{geoloc17}; and ownership attribution has
its own uncertainty~\cite{rocketfuel,bdrmapit}. Our audit complements
cartography. Maps can supply or validate explicit joins; the audit tests
shared-risk evidence at the granularity where each public source is
strongest.

\textbf{Behavioral localization and event forensics.}
Giotsas et al.\ localized interconnections to colocation
facilities~\cite{giotsas15} and later combined BGP signals with facility
metadata to localize peering-infrastructure outages~\cite{giotsas17},
demonstrating that building-level events can produce correlated
inter-domain failures. Internet outage systems and incident analyses
provide complementary event evidence~\cite{ripeegypt08,cho11,trinocular,ioda,weather19}.
Our temporal experiment starts from the same broad observation---shared
physical failures can leave correlated routing signatures---and tests a
specific 60-second predicate against controls before using it as
evidence.

\textbf{From public data to an audit.}
Routing archives~\cite{routeviews,ripe,bgpstream}, interconnection
registries~\cite{peeringdb}, cable atlases~\cite{telegeography}, and
measurement platforms~\cite{ark} now expose different parts of the
physical-risk problem. Industry discussions likewise identify physical
path independence as a resilience objective even when exact underlay
data remain private~\cite{cloudflare25,ofds26}. What is missing is a way
to combine these observations without turning an operator- or
cable-level signal into an unsupported end-to-end path claim. Our audit
keeps each signal at its native granularity and records every mapping
used to connect signals across layers.

\section{Terms and Metrics}
\label{sec:model}

\textbf{Paths and hidden resources.}
A path $P$ between endpoints $(s,t)$ is a logical sequence of ASes and IP
links carried by a physical resource set $R(P)\subset U$. The resource
universe includes conduit segments, facilities, landing zones, and power
domains. The logical sequence is observable; the complete $R(P)$
generally is not.

\textbf{Logical diversity.}
For a pair $A=(P_1,P_2)$ between the same endpoints, let
$\mathrm{LogDiv}(A)=1$ when the observed paths share neither an AS nor an
IP link. We define IRI only for such logically disjoint pairs, so the
physical question is not conflated with cases already non-disjoint at
Layer~3.

\textbf{Illusory Resilience Index.}
Let $w:U\rightarrow\mathbb{R}_{>0}$ assign a hazard weight to each
physical resource. For a logically disjoint pair, define
\begin{equation}
\mathrm{IRI}(A) \;=\;
\frac{\sum_{d\,\in\,R(P_1)\cap R(P_2)} w(d)}
     {\sum_{d\,\in\,R(P_1)\cup R(P_2)} w(d)}
\;\in\;[0,1].
\label{eq:iri}
\end{equation}
IRI is the hazard-weighted physical overlap hidden behind a logical
diversity claim. IRI $=0$ denotes physical independence under the chosen
resource universe and weights; larger values denote greater shared
exposure. The exact resource sets and weights needed to compute
Eq.~\ref{eq:iri} are normally unavailable outside the carriers. We
therefore use IRI as a \emph{latent target}, not an empirically
estimated scalar.

\textbf{What the audit observes.}
The public datasets observe different objects: routing archives expose
origin-group behavior, PeeringDB exposes operator facilities and
exchange memberships, and cable atlases expose cable systems and landing
points. A firing is evidence at that channel's native granularity. A
specific end-to-end path claim requires an explicit, independently
verifiable join from that native object to the path being audited.

We use three evidence states. \emph{No detected signal} means no
applicable discriminative channel fires; a \emph{single-channel
candidate} has one firing; and a \emph{corroborated cross-layer dossier}
has two or more discriminative firings on explicitly joined objects.
These labels are not probabilities and do not estimate IRI. A joint
path-level probability would require a matched joint-null experiment.


\section{Public-Data Audit}
\label{sec:methodology}

We apply three rules throughout the audit. First, each channel is judged
only on the object its source actually observes. Second, each firing
predicate is fixed before its control rate is measured. Third, we combine
signals across channels only when an explicit public mapping connects
their native objects. A cable-system observation is never promoted
silently into an operator- or path-level claim.

\subsection{Audited Objects and Datasets}
\label{sec:object}
\label{sec:datasets}

The latent target in Section~\ref{sec:model} is a logically disjoint
end-to-end path pair. The public sources expose three coarser native
objects:
\begin{itemize}[topsep=2pt,itemsep=1pt,parsep=0pt,leftmargin=*]
\item \emph{Temporal:} a pair of monitored origin groups during a
  documented physical incident.
\item \emph{Structural:} a pair of network operators and their listed
  facility/IXP footprints.
\item \emph{Geographic:} a pair of distinct submarine cable systems
  serving the same intercontinental corridor.
\end{itemize}

A firing therefore speaks only at its native granularity. Cross-layer
dossiers record every mapping used to connect objects, such as cable
consortium membership or operator ownership. A verified end-to-end path
claim requires a path-level join that these public sources do not
generally provide.

Table~\ref{tab:data} lists the live datasets used in the paper. All are
public and can be rerun as the underlying registries and archives change.
The BGP replay is the most expensive step: after candidate origins are
fixed, each incident window runs in minutes. The structural and
geographic analyses run in seconds to minutes on a workstation. We will
release the pipelines and per-figure extracts.

\begin{table}[t]
\centering
\caption{Public datasets and the native object audited by each empirical
channel. Every reported statistic derives from these snapshots.}
\label{tab:data}
\footnotesize
\begin{tabularx}{\columnwidth}{@{}p{1.35cm} p{1.65cm} L p{1.2cm}@{}}
\toprule
\textbf{Channel} & \textbf{Native object} & \textbf{Source / scope} &
\textbf{Retrieved} \\
\midrule
Temporal & Origin-group pair in an incident & RouteViews route-views2
and RIPE RIS rrc00 via PyBGPStream; $\pm30$\,min around the five 2024
incidents of Table~\ref{tab:events} & Aug.~5, 2026 \\
Structural & Operator pair & PeeringDB bulk dump; facilities and IXP
memberships for 47 operators in ten regions & Aug.~4, 2026 \\
Geographic & Cable-system pair & TeleGeography public GeoJSON; 68
systems, 606 landing points, twelve corridors & Aug.~4, 2026 \\
\bottomrule
\end{tabularx}
\vspace{-1.5em}
\end{table}

\subsection{Temporal Signal: Correlated BGP Withdrawals}
\label{sec:ch:temporal}

\textbf{Predicate.}
A common physical failure can affect routes associated with multiple
networks at roughly the same time. We test a simple version of this
signal. For a monitored origin-group pair, $f_T$ fires when both groups
show a withdrawal spike in the same $\Delta t=60$\,s bucket during a
documented physical incident, after the artifact screens below. A firing
is evidence of correlated routing failure; it does not localize the
failed resource or establish shared fiber by itself.

\textbf{Incident replay.}
We replay RouteViews and RIPE RIS update streams through PyBGPStream in a
$\pm30$-minute window around each event in a pre-specified catalog of
2024 physical incidents (Table~\ref{tab:events}), assembled from operator
reports, outage monitors~\cite{trinocular,ioda}, and press coverage.
Monitored origin groups are fixed from external facts alone (consortium
membership, landing parties, documented dependence), never from the
replay windows themselves; the group definitions and per-ASN
justifications ship with the artifact.
Three steps convert raw updates into a monitored series. \emph{Origin
attribution:} each withdrawal is assigned to the origin learned from
announcements in a two-hour warm-up. \emph{Artifact screening:} we
exclude a collector peer that withdraws more than 20{,}000 prefixes in
one minute or more than half of its learned table as a session reset; we
discard a minute whose withdrawals span more than five times the
window's median origin breadth as a collector-side event; and we assign
an AS-SET origin only when it maps to exactly one monitored group.
\emph{Spike detection:} withdrawals are bucketed per minute per origin
group, and a bucket is a spike when its count exceeds a robust median
$+\,k\cdot$MAD threshold learned from the window's quiet periphery.

Because the windows are seeded by documented incidents, the positive
statistic is a hit rate $\hat p_T$: the fraction of monitored
incident/group pairs for which $f_T$ fires. Section~\ref{sec:ch:calibration}
measures how often the same predicate fires on temporal controls and
within-window permutations.

\subsection{Structural Signal: Interconnection Overlap}
\label{sec:ch:structural}

\textbf{Predicate.}
Distinct candidate operators can still share substantial
interconnection infrastructure. We score each operator pair by the
Jaccard similarity of their listed PeeringDB \emph{facility} sets,
\[
J = \frac{|F_1\cap F_2|}{|F_1\cup F_2|},
\]
so the predicate quantity is site-level by construction. The structural
predicate $f_S$ fires when $J\ge\tau=0.20$, with $\tau$ fixed before
calibration.

A shared street address is a shared exposure to building-level events,
which is what gives the score its physical interpretation; it still does
not claim that a particular flow traverses that building. IXP
co-membership is coarser, since two members need not occupy the same
site, so exchanges are reported alongside the score rather than inside
it, and path-level claims are reserved for cases with an explicit
router/facility join.

\textbf{Pipeline.}
We analyze a curated set of 47 major transit, content, and incumbent
operators spanning ten world regions. The set includes plausible
alternatives available to multihomed customers in each region, and the
artifact contains the exact list. For every operator pair, we compute the facility Jaccard score, recording IXP co-membership alongside it. We repeat the headline comparison after
restricting the sample to operators that list at least five facilities,
which tests sensitivity to sparse self-reporting. We also rank individual
facility addresses by the number of operator pairs colocated there.

\subsection{Geographic Signal: Shared Landing Zones}
\label{sec:ch:geographic}

\textbf{Predicate.}
Distinct submarine systems can converge at the same coastal gateway,
creating a shared shore-end hazard even when their mid-ocean routes
separate. For a cable-system pair $(C_1,C_2)$ serving the same
intercontinental corridor, $f_G$ fires when the systems have landing
points inside a common coastal hazard zone of radius $r=25$\,km. The
radius is an operational neighborhood, not a universal damage radius;
Appendix~\ref{app:sensitivity} repeats every corridor calculation for
$r\in\{10,25,50\}$\,km.

\textbf{Pipeline.}
We ingest TeleGeography's public GeoJSON cable geometries and landing
points, match landing stations to cable endpoints, and test every
within-corridor pair. A corridor is a named intercontinental route, such
as Europe--Asia via the Red Sea, with a fixed list of cable systems from
TeleGeography's route classification. The artifact contains all twelve
corridor definitions. The study covers 68 cable systems and 606 landing
points.

TeleGeography's published routes are stylized rather than surveyed
mid-ocean geometry, so all geographic claims remain at landing-station
granularity. That is also the relevant scale for localized shore-end
cuts, anchor damage, and single-gateway geographies.

\subsection{Calibration: Channel-Specific Null Models}
\label{sec:ch:calibration}

A signal is useful only if its predicate is uncommon under a suitable
control construction. For each channel we estimate a \emph{null firing
rate} $q_c$: the firing rate of the fixed predicate in a reference
population where the target dependence is less plausible. Where
documented incidents provide labeled positives, we also report a hit
rate $p_c$. These are calibration controls, not certified ground-truth
negatives.

\begin{itemize}[topsep=2pt,itemsep=1pt,parsep=0pt,leftmargin=*]
\item \emph{Temporal:} we use (i) control origin groups on other
continents with disjoint upstreams, replayed through the same windows,
and (ii) a within-window permutation that circularly shifts one group's
spike series and recounts same-bucket coincidences. We use the larger
null estimate.
\item \emph{Structural:} the primary control population is all 976
inter-region operator pairs at the declared $\tau$. A supplementary
size-matched check rewires the operator--facility/IXP bipartite graph
while preserving node degrees (1{,}000 graphs).
\item \emph{Geographic:} the control population is all 2{,}625
cross-corridor cable pairs. This is conservative because some
cross-corridor systems genuinely share hub landings, notably Singapore
and the Egyptian coast, so genuine shared exposure is counted inside the
null. Documented same-event cable damage in Appendix~\ref{app:incidents}
supplies labeled positives for $p_G$.
\end{itemize}

Table~\ref{tab:calibration} reports all rates with 95\% Wilson intervals.
The calibrations are marginal: structural and geographic signals use
different native objects and control populations. We therefore do not
multiply their marginal null rates into a joint path-level probability.
Cross-channel evidence is combined only through explicit dossiers.

\subsection{Cross-Channel Dossiers}
\label{sec:fusion}

For each practical diversity question, a dossier records (i) the native
object tested by each channel, (ii) the result of its fixed predicate,
and (iii) the public mapping that connects objects across channels. We
use three labels:
\begin{itemize}[topsep=2pt,itemsep=1pt,parsep=0pt,leftmargin=*]
\item \emph{No detected signal}: no applicable discriminative channel
fires.
\item \emph{Single-channel candidate}: one discriminative channel fires
at its native granularity.
\item \emph{Corroborated cross-layer dossier}: at least two
discriminative channels fire on objects connected by an explicit public
mapping.
\end{itemize}
These labels separate one public warning sign from cases where distinct
sources point to the same shared-risk concern. They do not estimate IRI
or certify end-to-end path overlap.

\section{Empirical Results}
\label{sec:results}

Two of the three empirical signals survive calibration. The structural
and geographic analyses use live Aug.~4, 2026 snapshots; the temporal
experiment replays six windows around five documented 2024 incidents
using archives retrieved Aug.~5, 2026. Table~\ref{tab:results} summarizes
the results and the native object supported by each source.

\begin{table*}[t]
\centering
\caption{Summary of empirical results. Each row reports the native
object actually supported by the public source.}
\label{tab:results}
\footnotesize
\begin{tabularx}{\textwidth}{@{}p{2.5cm} p{3.1cm} L p{3.7cm}@{}}
\toprule
\textbf{Signal} & \textbf{Scope} & \textbf{Headline result} &
\textbf{Interpretation} \\
\midrule
Temporal: BGP co-withdrawal & Five 2024 incidents; 12 monitored
positive pairs; RouteViews + RIS & No same-minute hit (0/12, CI
[0, 0.243]); permutation null 0.444 & The 60\,s predicate is rejected
as non-discriminative and is not used for corroboration \\
Structural: interconnection footprint & 47 operators, ten regions,
1{,}081 pairs; PeeringDB & Intra-region facility-Jaccard mean 0.094 (median
0.067) vs.\ inter-region 0.024 (median 0); gap persists at $\ge5$
facilities; one facility colocates 153 pairs & Operator-level
interconnection concentration; facility ranking identifies site-level
shared exposure \\
Geographic: landing zones & Twelve corridors, 68 systems, 606 landings;
TeleGeography & Within-corridor convergence 31--100\%; two corridors
saturated; 25\,km rule recovers 10/11 documented co-damaged pairs &
Cable-system shared shore-end exposure and geographic diversity ceiling \\
Cross-channel dossiers & Explicit mappings between audited objects &
Separate structural and geographic signals can corroborate a shared-risk
concern when the mapping is explicit & Qualitative cross-layer
corroboration; no joint probability or IRI estimate is assigned \\
\bottomrule
\end{tabularx}
\end{table*}

\subsection{Temporal: Co-Withdrawal Fails Calibration}
\label{sec:res:temporal}

The 60-second temporal predicate fails its calibration test. Across the
twelve monitored positive pairs, $f_T$ produces no same-bucket
co-withdrawal ($\hat p_T=0/12$, 95\% CI $[0.000,0.243]$). The permutation
control produces a same-bucket coincidence in 44.4\% of shuffles
($\hat q_T=0.444$ $[0.429,0.459]$, 24{,}000 trials). At these public
collectors and this granularity, minute-level co-withdrawal is not a
discriminative shared-failure signal.

\begin{figure*}
\centering
\includegraphics[width=\textwidth]{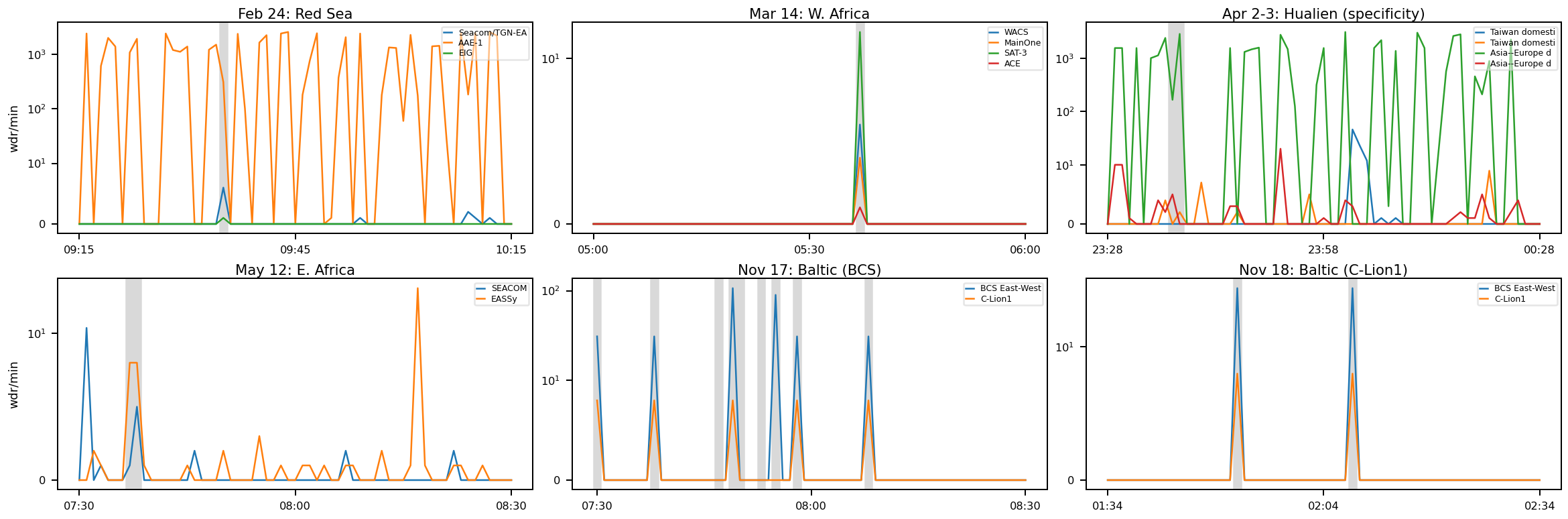}
\caption{Temporal experiment: origin-attributed withdrawal series for
six $\pm30$-minute incident windows (symlog; shaded minutes are
collector artifacts removed by the breadth screen). Post-cut behavior is
churn and staggered episodes rather than a clean same-minute signature.}
\label{fig:year}
\end{figure*}

The replay explains why. We analyzed six $\pm30$-minute windows (one for
each of the two Baltic damages), using both collectors and full origin
attribution. The Red Sea hour alone contains 648{,}101 withdrawals across
more than 23{,}000 origins. The artifact screens remove a one-minute
collector-side event in which 22{,}489 origins withdraw at once, sixty
times the median breadth, and ten (peer, minute) session-reset buckets.
Fig.~\ref{fig:year} shows the resulting monitored series. Post-cut routing
behavior appears as sustained churn and staggered episodes: the East
African groups rise in different minutes, the largest AAE-1 dependent
withdraws throughout the Red Sea hour, and the Baltic damages occur
roughly eighteen hours apart.

We therefore exclude the 60-second predicate from the current dossiers.
This result is also a check on the audit itself: calibration removes a
plausible but noisy signal instead of allowing it to influence the
conclusions. Episode-level temporal predicates are the next experiment
(Sec.~\ref{sec:future}).

\subsection{Structural: Interconnection Concentration}
\label{sec:res:structural}

\begin{figure}
\centering
\includegraphics[width=0.98\columnwidth]{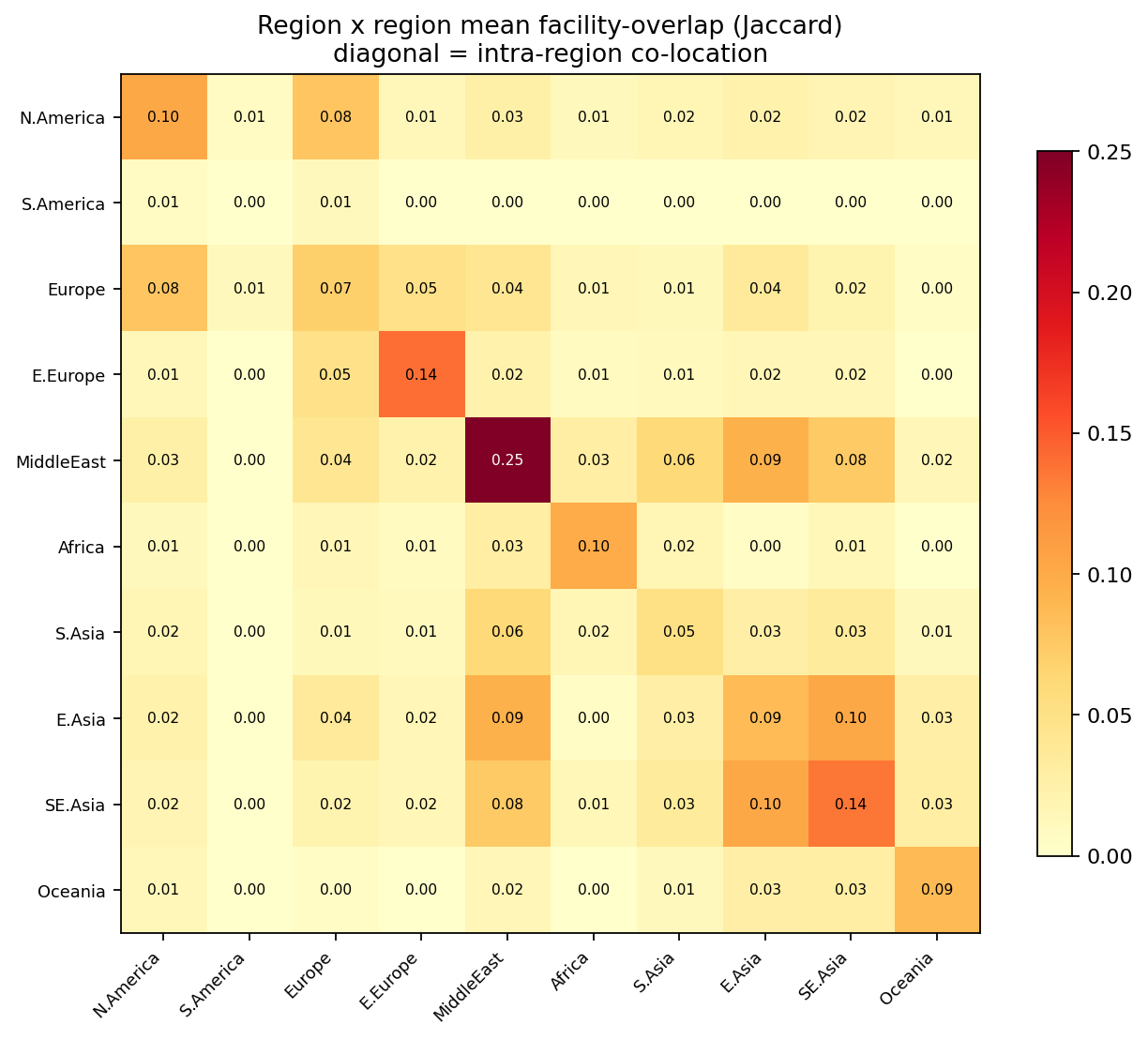}
\caption{Structural signal: mean facility-footprint Jaccard score by region pair (diagonal = same region), from the live PeeringDB
snapshot.}
\label{fig:regionmatrix}
\end{figure}

Same-region operator pairs have substantially greater listed
interconnection-footprint overlap than cross-region pairs
(Fig.~\ref{fig:regionmatrix}). Across 1{,}081 pairs, mean same-region
facility overlap is $0.094$ (median $0.067$), compared with an
inter-region mean of $0.024$ and a median of zero. A typical
same-region candidate alternative in this sample shares some listed
interconnection footprint; a typical cross-region pair shares none.

The gap survives a simple reporting-density check. Restricting the sample
to operators that list at least five facilities gives means of $0.118$
and $0.030$, again a $3.9\times$ ratio. The pattern is also
region-specific: the Middle East has the largest same-region mean
($0.25$). SEACOM and Liquid Intelligent Technologies are the highest-scoring
pair at $J=0.42$ (14 shared facilities; the two networks also share 13
exchanges), and they can plausibly appear as alternatives in East
Africa. Folding IXP memberships into the score as typed elements does
not change the picture: the same-region mean becomes $0.106$ (median
$0.090$) against $0.036$ (median $0.016$) cross-region, and the
SEACOM--Liquid pair remains the maximum, rising slightly to $0.46$.

\begin{figure}
\centering
\includegraphics[width=0.98\columnwidth]{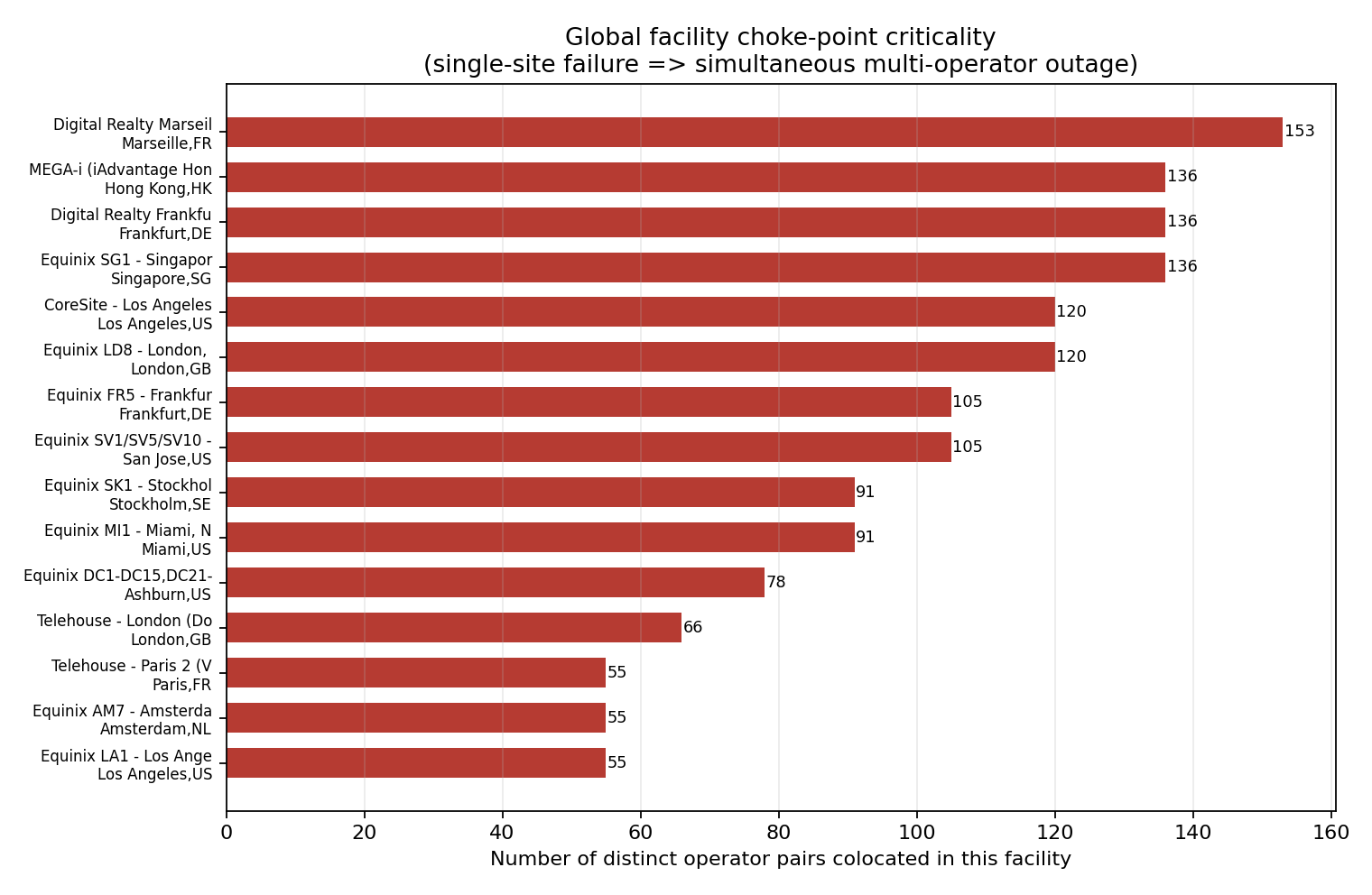}
\caption{Facility criticality: operator pairs colocated at each listed
facility address. A site-level failure exposes the listed pairs to a
common physical event; the figure does not assert that every circuit
traverses that facility.}
\label{fig:faccrit}
\vspace{-0.5em}
\end{figure}

The facility-only ranking gives this footprint concentration a direct
site-level interpretation (Fig.~\ref{fig:faccrit}). The Digital Realty
Marseille campus contains 153 operator pairs from our sample; MEGA-i
(Hong Kong), Digital Realty Frankfurt, and Equinix SG1 contain 136 each;
One Wilshire and Equinix LD8 contain 120 each. A power, fire, or flood
failure at any of these sites can expose many candidate operator pairs to
the same physical event.

\subsection{Geographic: Corridor Diversity Collapse}
\label{sec:res:geographic}

\begin{figure}
\centering
\includegraphics[width=0.98\columnwidth]{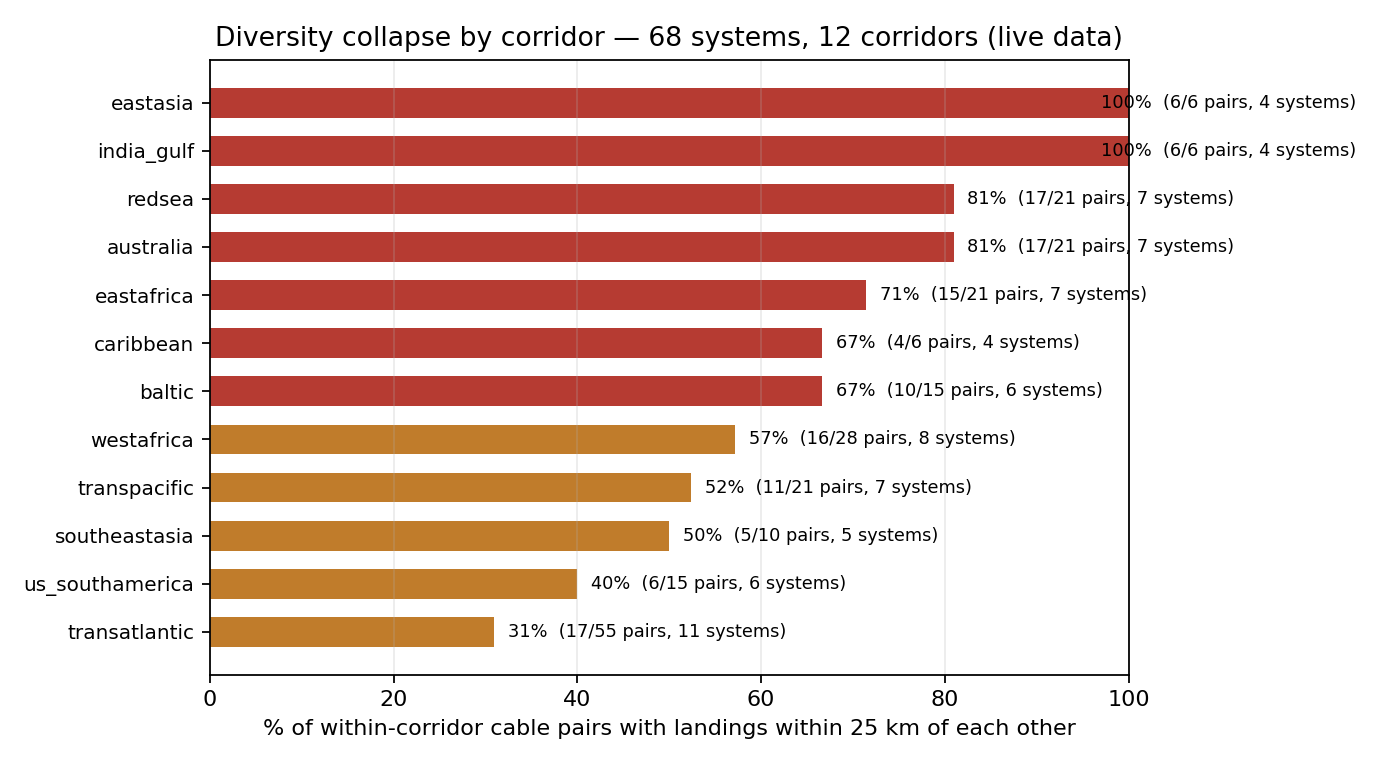}
\caption{Geographic signal: share of distinct within-corridor
cable-system pairs with landings inside a common $r=25$\,km coastal
hazard zone (live TeleGeography data).}
\label{fig:collapse}
\end{figure}

Within-corridor landing convergence spans 31--100\% at a 25\,km hazard
radius (Fig.~\ref{fig:collapse}). East Asia and India--Gulf are saturated
(100\%, 6/6 pairs each); the Red Sea and Australia reach 81\%, East
Africa 71\%, the Caribbean and Baltic 67\%, West Africa 57\%, the
transpacific 52\%, Southeast Asia 50\%, US--South America 40\%, and the
transatlantic 31\% (17/55).

The ordering is robust to the operational radius. Across
$r\in\{10,25,50\}$\,km, no corridor moves by more than 20 percentage
points; the two saturated corridors and the 31\% transatlantic value are
invariant (Appendix~\ref{app:sensitivity}). In single-gateway
geographies, this convergence sets a physical ceiling on the shore-end
diversity available from the cable systems in our corridor set.

The global union of all 606 landing points identifies the largest coastal
choke clusters (Fig.~\ref{fig:chokemap}, Table~\ref{tab:chokeclusters}).
Tuas and Chikura each place eleven distinct cable systems inside one
25\,km zone; Jeddah, Chung Hom Kok, and Shima place eight. The 2024
incident record shows the same geography: several systems damaged in the
Red Sea events use the Jeddah and Egyptian landing clusters.

\begin{table}[t]
\centering
\caption{Largest coastal choke clusters at $r=25$\,km: distinct cable
systems landing within one hazard zone.}
\label{tab:chokeclusters}
\footnotesize
\begin{tabularx}{\columnwidth}{@{}L r@{}}
\toprule
\textbf{Cluster} & \textbf{Systems} \\
\midrule
Tuas, Singapore & 11 \\
Chikura, Japan & 11 \\
Jeddah, Saudi Arabia & 8 \\
Chung Hom Kok, Hong Kong & 8 \\
Shima, Japan & 8 \\
Sesimbra, Portugal & 7 \\
\bottomrule
\end{tabularx}
\vspace{-1em}
\end{table}

\subsection{Calibration Results}
\label{sec:res:calibration}

Calibration leaves two discriminative signals. The structural predicate
at $\tau=0.20$ fires on 15 of 976 inter-region control pairs
($\hat q_S=0.015$ [0.009, 0.025]). The geographic predicate fires on
10.3\% of cross-corridor control pairs ($\hat q_G=0.103$ [0.092, 0.115]).
Because several cross-corridor pairs share real hub landings, the latter
control intentionally counts some genuine shared exposure inside its
null. Against documented co-damage, the same 25\,km rule recovers 10 of
11 positive cable pairs ($\hat p_G=0.909$ [0.623, 0.984]). The one miss
is C-Lion1 / BCS East--West: the systems were damaged in the same
multi-day Baltic event, but their landing points are 357\,km apart, so a
landing-zone predicate should remain silent (Appendix~\ref{app:incidents}).

The temporal predicate is rejected: its null firing rate is 0.444 and
its positive hit rate is 0/12. Table~\ref{tab:calibration} reports the
marginal calibration results for all three channels.

\begin{table}[t]
\centering
\caption{Marginal channel calibration. $\hat q_c$ is the firing rate
under the stated control construction; $\hat p_c$ is the hit rate on
documented positives (95\% Wilson intervals).}
\label{tab:calibration}
\footnotesize
\begin{tabularx}{\columnwidth}{@{}l r l r l@{}}
\toprule
\textbf{Channel} & $n_{\mathrm{ctrl}}$ & $\hat q_c$ [CI] &
$n_{\mathrm{pos}}$ & $\hat p_c$ [CI] \\
\midrule
Temporal & 24{,}000 & 0.444 [0.429,\,0.459] & 12 & 0.000 [0.000,\,0.243] \\
Structural & 976 & 0.015 [0.009,\,0.025] & --- & --- \\
Geographic & 2{,}625 & 0.103 [0.092,\,0.115] & 11 & 0.909 [0.623,\,0.984] \\
\bottomrule
\end{tabularx}
\vspace{-1em}
\end{table}

\subsection{Cross-Channel Dossiers}
\label{sec:res:dossiers}

Structural and geographic measurements use different random units:
operator pairs and cable-system pairs. We combine them only when a public
mapping connects those objects, and we assign no numerical joint
probability to the resulting dossier.

\emph{Case 1: SEACOM / EASSy (East Africa), corroborated cross-layer
dossier.}
The geographic predicate fires: SEACOM and EASSy co-land at Mombasa,
Djibouti, and Mtunzini, and the pair is one of the documented positives
recovered by $f_G$ after the May~12 co-damage near
Mtunzini~\cite{eastafrica24} (Appendix~\ref{app:incidents}). At operator
granularity, Liquid lists itself as an EASSy consortium
member~\cite{liquidnetwork26}; the SEACOM--Liquid pair is the strongest
structural pair in the sample at $J=0.42$. The explicit cable/operator mapping therefore connects two distinct
public signals: at their native granularities, this pair is disjoint in
name only. That is corroborated evidence across those objects, not a
verified end-to-end circuit overlap.

\emph{Case 2: SEA-ME-WE-5 / AAE-1 (Europe--Asia), single-channel
candidate.}
The geographic predicate fires at the Abu Talat and Jeddah landing zones,
and the pair appears among the documented Red Sea co-damage positives.
The curated structural set has no one-to-one operator mapping for the two
cable consortia, so this remains a geographic candidate rather than a
cross-layer corroborated case.

\emph{Case 3: Hualien, construction sanity check.}
The Apr.~3 earthquake caused substantial domestic disruption while
international cables were largely unaffected~\cite{taiwan24}. The event
catalog contains no positive international cable pair for this window,
and the monitored international groups produce only one same-bucket
coincidence, at exactly the chance level the $0.444$ null predicts
(Table~\ref{tab:events}). Silence is not proof of independence; the
case shows that the incident construction neither forces a cross-system
positive when the documented record contains none, nor hides the chance
floor when a coincidence does appear.

The dossiers make the audit output explicit: what each public source
establishes, how objects are joined across layers, and where path-level
verification is still required.

\begin{strip}
\centering
\includegraphics[width=0.88\textwidth]{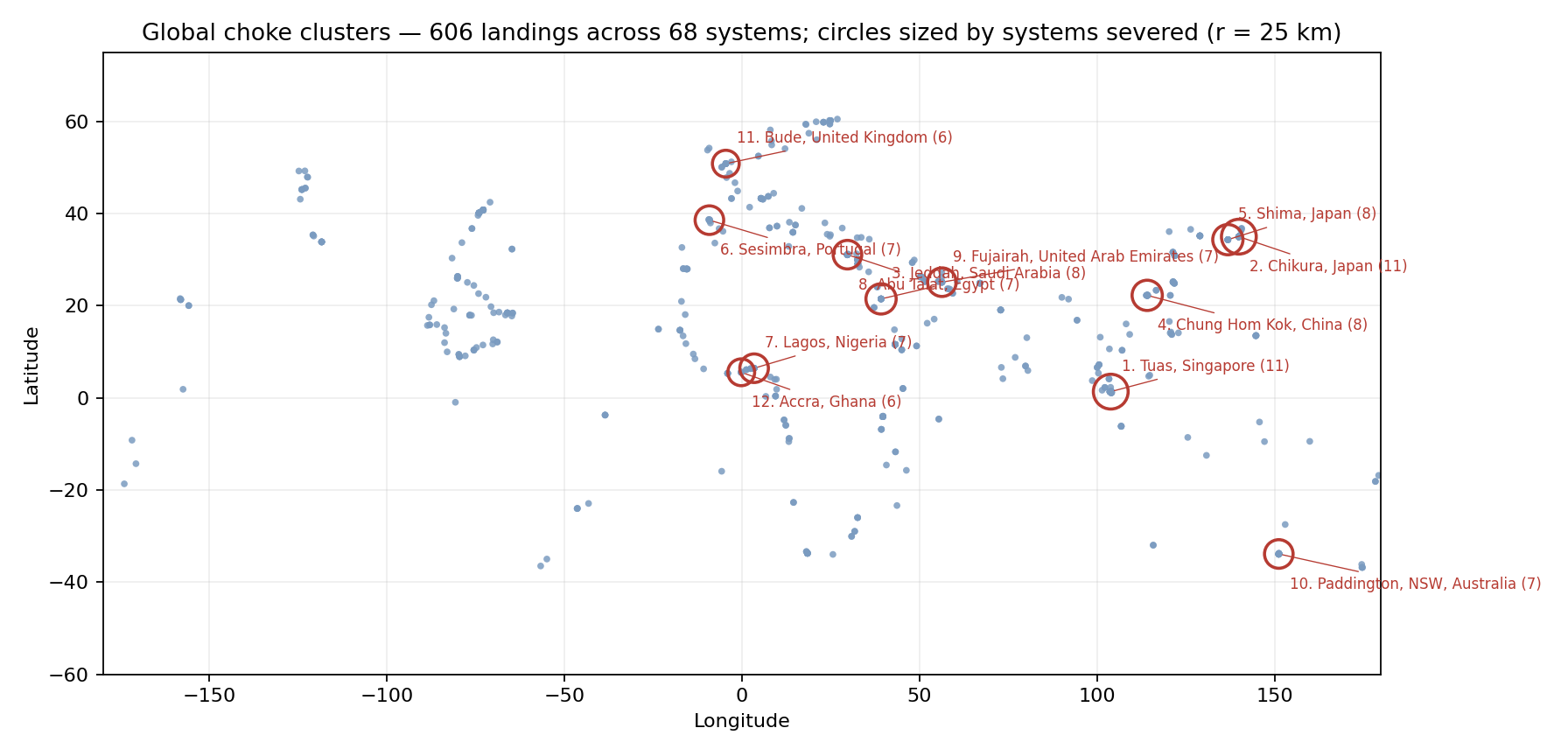}
\captionof{figure}{Global coastal choke clusters at $r=25$\,km. Marker
size encodes the number of distinct cable systems landing inside one
hazard zone (Table~\ref{tab:chokeclusters}).}
\label{fig:chokemap}
\end{strip}


\section{Discussion}
\label{sec:discussion}

The measurements support four conclusions. First, shared-risk exposure is
strongly geographic: same-region operator pairs have materially more
overlap in their listed interconnection footprints than cross-region pairs.
Second, coastal geography sets a hard ceiling on shore-end submarine
diversity in several corridors, including two saturated corridors in
our sample. Third, intuitive behavioral signals need calibration: the
60-second BGP co-withdrawal predicate fails its test. Fourth,
public-data audits are most reliable when each observation remains at
its native granularity instead of being converted into unsupported
path-level precision.

\subsection{Operational Use}
\label{sec:leading}

Operators already treat physical path independence as a resilience
objective even when exact ground truth is private~\cite{cloudflare25};
post-disaster reviews likewise call for better disclosure of fiber
infrastructure~\cite{ofds26}. The audit identifies where stronger
evidence is worth requesting.

For procurement, a structural or geographic firing marks alternatives
that merit carrier attestation before being treated as independent. For
restoration planning, facility and landing-zone criticality identify
sites where several nominal alternatives share exposure. For
inter-provider cooperation, the important question is whether the backup
has a different dependency footprint, not merely a different logical
route. The audit does not replace carrier ground truth; it identifies
where that ground truth matters most.

\subsection{Extending the Audit with Latency--Geometry}
\label{sec:latencyextension}

A fourth channel can extend coverage to terrestrial long-haul paths,
where facility records and cable landings provide less information. The
basic observation is geometric: light in single-mode fiber travels at
roughly $204{,}000$\,km/s, so one millisecond of round-trip time
corresponds to about 102\,km of one-way fiber. Repeated minimum-RTT
measurements can therefore bound path length and circuitousness~\cite{singla14}.

Our specified protocol uses the RIPE Atlas anchoring mesh (1{,}771
registered anchors with operator-declared coordinates). It constructs
co-endpoint anchor pairs between the same two metros in different
networks, verifies logical disjointness of the observed paths, converts
minimum RTT to implied fiber length, and tests whether two candidate
paths agree within a tolerance $\varepsilon$. Calibration uses
different-endpoint path pairs matched on great-circle distance as the
control population. We sweep $\varepsilon\in\{0.2,0.5,1.0\}$\,ms and
select the largest value whose control firing rate is at most 0.20.
Equal path length is only a screen: parallel but physically distinct
rights-of-way can have similar lengths. This channel should enter the
audit only if its live calibration demonstrates discrimination, so it is
outside present empirical evaluation.

\subsection{Future Measurement}
\label{sec:future}

Three extensions follow directly from the results. A matched
cross-channel control set can measure joint structural and geographic
co-firing directly. Traceroute-to-facility mapping and verified
operator--cable--path joins can move dossiers from operator/cable objects
toward specific end-to-end circuits~\cite{geoloc17,bdrmapit}. Finally,
the temporal data should be tested at episode rather than 60-second
granularity: the live replays show sustained and staggered post-failure
behavior that a one-minute predicate misses. The pipelines are
incremental, so each extension can be evaluated as new registry
snapshots and incident evidence become available.

\section{Limitations}
\label{sec:limitations}

\textbf{Granularity and proxy scope.}
IRI is a path-level latent quantity, while the empirical channels observe
origin groups, operator footprints, and cable landing points. A shared
facility address establishes common site-level exposure but does not
show that a particular circuit traverses that facility; IXP
co-membership is coarser still. Likewise, a shared landing zone
establishes shore-end exposure, not mid-ocean route overlap. The dossiers
preserve these distinctions and require an explicit mapping between
native objects. They do not certify end-to-end physical overlap or estimate IRI
magnitude.

\textbf{Coverage and control constructions.}
PeeringDB is self-reported and uneven in density. The same-region versus
cross-region gap persists after restricting the sample to operators that
list at least five facilities, but this cannot remove reporting bias in
principle. TeleGeography's public geometries are schematic, so we keep
geographic claims at landing-station granularity. The study covers 47
curated operators, twelve corridors, and documented incidents from one
year; it is geographically broad but not exhaustive. The inter-region
and cross-corridor calibration populations are reference controls, not
guaranteed physical negatives. We therefore report their observed rates
as \emph{null firing rates}, not universal false-positive probabilities.

\textbf{Calibration scope.}
Structural and geographic null rates are measured marginally on
different native objects, so the paper assigns no joint probability to
cross-channel corroboration. Such a probability requires a matched
joint-null experiment. Conversely, silence means only that the
applicable public instruments detected no shared-risk signal; it does
not certify IRI $=0$.

\textbf{Ethics.}
All inputs are public: archived routing updates, self-published
interconnection registries, a public cable atlas, and documented
infrastructure incidents. The study generates no probing load, touches
no user traffic or personal data, and reports only infrastructure
relationships already visible in public sources. The released pipeline
allows every firing and aggregate statistic to be re-derived.

\section{Conclusion}
\label{sec:conclusion}

Logical disjointness is easy to observe; physical independence is not.
We define illusory resilience as this gap and IRI as the latent
hazard-weighted overlap of a logically disjoint path pair. Empirical
results show that useful parts of this hidden risk are visible without a
complete underlay map.

Across 47 operators, same-region pairs have substantially greater listed
interconnection-footprint overlap than cross-region pairs, and
individual facilities colocate more than 150 candidate operator pairs.
Across 68 submarine systems in twelve corridors, 31--100\% of
within-corridor pairs share a 25\,km coastal hazard zone; two corridors
are saturated, and the same geographic predicate recovers ten of eleven
cable pairs documented as co-damaged in 2024. The BGP experiment adds an
important negative result: minute-level co-withdrawal is too noisy to
serve as a shared-failure signal at the tested collectors. Calibration excludes it from the audit.

A backup can therefore be disjoint in the routing graph and still share
consequential physical exposure with the primary. Public data cannot
certify every hidden resource on an end-to-end path, but it can identify
where nominal diversity requires stronger physical-path verification.

\bibliographystyle{unsrt}
\bibliography{refs}
\clearpage
\appendix

\twocolumn[
\begin{center}
{\Large\bfseries APPENDIX}
\end{center}

\vspace{0.5em}

\begin{minipage}{\textwidth}
\centering

\captionof{table}{Replay results for documented 2024 incidents
using the pre-specified 60\,s co-withdrawal predicate.}
\label{tab:events}

\vspace{-0.4em}

\footnotesize
\setlength{\tabcolsep}{3pt}
\renewcommand{\arraystretch}{0.90}

\begin{tabularx}{\textwidth}{
    @{}
    p{1.55cm}
    L
    p{2.0cm}
    p{1.55cm}
    p{2.15cm}
    @{}
}
\toprule
\textbf{Window} &
\textbf{Documented incident} &
\textbf{Source} &
\textbf{Monitored groups} &
\textbf{Co-failure?} \\
\midrule
Feb.~24 &
Red Sea cable damage involving Seacom/TGN-EA, AAE-1, and EIG &
\cite{redsea24} &
3 monitored &
None within 60\,s \\

Mar.~14 &
West African cuts off C\^ote d'Ivoire involving WACS, MainOne,
SAT-3, and ACE &
\cite{safrica24} &
4 monitored &
None within 60\,s \\

Apr.~3 &
Hualien earthquake ($M_w$ 7.4); domestic disruption, with
international cables largely unaffected &
\cite{taiwan24} &
4 monitored &
One pair; chance-level \\

May~12 &
East African cuts near Mtunzini involving SEACOM and EASSy &
\cite{eastafrica24} &
2 monitored &
Elevated in different minutes \\

Nov.~17--18 &
Baltic incidents involving BCS East--West and C-Lion1 &
\cite{baltic24} &
$2\times2$; two windows &
None; damages ${\sim}18$\,h apart \\
\bottomrule
\end{tabularx}

\vspace{1pt}

\begin{minipage}{0.98\textwidth}
\footnotesize
\emph{Note:} The Apr.~3 window is a specificity check.
International cable systems were largely unaffected, so
cross-system silence is the expected outcome.
\end{minipage}

\end{minipage}

\vspace{1.5em}
]

\section{Incident Sources, Labeled Positives}
\label{app:incidents}

Table~\ref{tab:positives} lists eleven cable pairs documented as damaged in the same physical event during 2024, the labeled positives behind $\hat{p}_G$ in Table~\ref{tab:calibration}, with minimum distance between two systems' landing points and the resulting predicate outcome at $r = 25$\,km. Four incident details matter for interpreting these labels. (1) \emph{Red Sea, Feb.~24}: three systems (Seacom/TGN-EA, AAE-1, EIG) were cut the same morning; EIG was already degraded from a December cut, which is relevant when interpreting its withdrawal series in the temporal replay~\cite{redsea24}. (2) \emph{West Africa, Mar.~14}: four systems failed nearly simultaneously off C\^ote d'Ivoire, a convergence point for all four routes; the suspected cause is a subsea rock slide~\cite{safrica24}. (3) \emph{East Africa, May~12}: SEACOM and EASSy failed together near Mtunzini, South Africa, which is within their shared landing corridor; the co-damage location is the predicted choke point~\cite{eastafrica24}. (4) \emph{Baltic, Nov.~17--18}: BCS East--West and C-Lion1 were damaged one day apart by a suspected dragged anchor along a track that crosses both routes; their landings lie 357\,km apart, so the landing-zone predicate is correctly silent, and this pair is the one miss in $\hat{p}_G$~\cite{baltic24}. The Apr.~3 Hualien earthquake appears in Table~\ref{tab:events} as a specificity check and contributes no positives: international cable systems were largely unaffected~\cite{taiwan24}.

\begin{center}
\begin{minipage}{0.98\columnwidth}
\centering

\captionof{table}{Documented 2024 co-damaged cable pairs used as
labeled positives for $\hat{p}_G$.}
\label{tab:positives}


\footnotesize
\setlength{\tabcolsep}{2.5pt}
\renewcommand{\arraystretch}{0.90}

\begin{tabularx}{\columnwidth}{@{}l L r l@{}}
\toprule
\textbf{Event} & \textbf{Cable pair} & \textbf{km} & $f_G$ \\
\midrule
Red Sea & SMW-5 / AAE-1 & 0.0 & fires \\
Red Sea & SMW-5 / EIG & 0.0 & fires \\
Red Sea & AAE-1 / EIG & 0.0 & fires \\
W.\ Africa & WACS / MainOne & 0.0 & fires \\
W.\ Africa & WACS / SAT-3 & 0.0 & fires \\
W.\ Africa & WACS / ACE & 0.0 & fires \\
W.\ Africa & MainOne / SAT-3 & 0.0 & fires \\
W.\ Africa & MainOne / ACE & 0.0 & fires \\
W.\ Africa & SAT-3 / ACE & 0.0 & fires \\
E.\ Africa & SEACOM / EASSy & 0.0 & fires \\
Baltic & C-Lion1 / BCS E--W & 357.2 & silent \\
\bottomrule
\end{tabularx}

\end{minipage}
\end{center}

\section{Calibration and Sensitivity Detail}
\label{app:sensitivity}

Table~\ref{tab:rsens} reports the radius sensitivity of the corridor-collapse percentages, computed on the live TeleGeography snapshot. The saturated corridors and the transatlantic floor are invariant across $r \in \{10, 25, 50\}$\,km; the maximum shift for any corridor is 20 points, confined to mid-table corridors.

\begin{center}
\begin{minipage}{0.98\columnwidth}
\centering

\captionof{table}{Radius sensitivity of corridor diversity
collapse on the live TeleGeography snapshot (\%).}
\label{tab:rsens}


\footnotesize
\setlength{\tabcolsep}{2.5pt}
\renewcommand{\arraystretch}{0.90}

\begin{tabular}{@{}l r r r r@{}}
\toprule
\textbf{Corridor} &
$r{=}10$ &
$r{=}25$ &
$r{=}50$ &
\textbf{Shift} \\
\midrule
East Asia & 100 & 100 & 100 & 0 \\
India--Gulf & 100 & 100 & 100 & 0 \\
Red Sea & 81 & 81 & 81 & 0 \\
Australia & 81 & 81 & 81 & 0 \\
East Africa & 71 & 71 & 71 & 0 \\
Caribbean & 67 & 67 & 83 & 16 \\
Baltic & 53 & 67 & 67 & 14 \\
West Africa & 57 & 57 & 57 & 0 \\
Transpacific & 52 & 52 & 52 & 0 \\
Southeast Asia & 40 & 50 & 70 & 20 \\
US--S.\ America & 27 & 40 & 40 & 13 \\
Transatlantic & 31 & 31 & 31 & 0 \\
\bottomrule
\end{tabular}

\end{minipage}
\end{center}

\end{document}